\documentclass{article}
\usepackage{ijcai17}

\usepackage{times}

\usepackage{url}

\usepackage{amsmath}
\usepackage{mathtools}
\newcommand*{\horzbar}{\rule[.5ex]{2.5ex}{0.5pt}}
\DeclareMathOperator*{\argmax}{arg\,max}

\title{Recognizing Abnormal Heart Sounds Using Deep Learning}
\author{Jonathan Rubin$^1$, Rui Abreu$^2$, Anurag Ganguli$^2$, Saigopal Nelaturi$^2$, Ion Matei$^2$, Kumar Sricharan$^2$\\ 
$^1$~Philips Research North America, $^2$~PARC, A Xerox Company\\
jonathan.rubin@philips.com, rui@computer.org,\\\{anurag.ganguli, saigopal.nelaturi, ion.matei, sricharan.kumar\}@parc.com}

\begin{document}

\maketitle

\begin{abstract}

The work presented here applies deep learning to the task of automated cardiac auscultation, i.e.~recognizing abnormalities in heart sounds. We describe an automated heart sound classification algorithm that combines the use of time-frequency heat map representations with a deep convolutional neural network (CNN). Given the cost-sensitive nature of misclassification, our CNN architecture is trained using a modified loss function that directly optimizes the trade-off between sensitivity and specificity. We evaluated our algorithm at the 2016 PhysioNet Computing in Cardiology challenge where the objective was to accurately classify normal and abnormal heart sounds from single, short, potentially noisy recordings. Our entry to the challenge achieved a final specificity of 0.95, sensitivity of 0.73 and overall score of 0.84. We achieved the greatest specificity score out of all challenge entries and, using just a single CNN, our algorithm differed in overall score by only 0.02 compared to the top place finisher, which used an ensemble approach.

\end{abstract}

\section{Introduction}
\label{sec:introduction}

Advances in deep learning \cite{lecun2015deep} are being made at a rapid pace, in part due to challenges such as ILSVRC -- the ImageNet Large-Scale Visual Recognition Challenge \cite{ILSVRC15}. Successive improvements in deep neural network architectures have resulted in computer vision systems that are better able to recognize and classify objects in images \cite{DBLP:journals/corr/LinCY13,DBLP:journals/corr/SzegedyVISW15} and winning ILSVRC entries \cite{DBLP:journals/corr/SzegedyLJSRAEVR14,DBLP:journals/corr/HeZRS15}. While a large focus of deep learning has been on automated analysis of image and text data, advances are also increasingly being seen in areas that require processing other input modalities. One such area is the medical domain where inputs into a deep learning system could be physiologic time series data. An increasing number of large scale challenges in the medical domain, such as \cite{web/kaggle/seizure-prediction} and \cite{web/kaggle/grasp-and-lift} have also resulted in improvements to deep learning architectures \cite{DBLP:conf/cvpr/LiangH15}.

\begin{figure}[h]
\includegraphics[width=1.0\columnwidth]{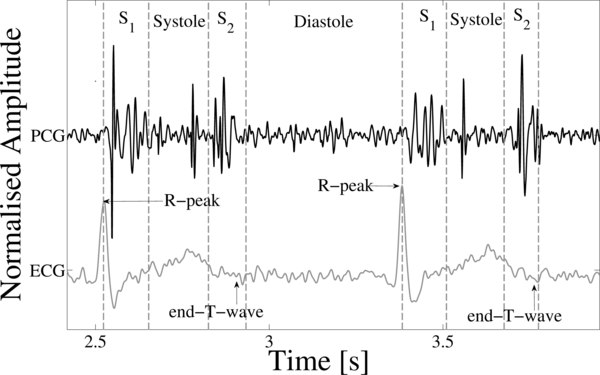}
\caption{A phonocardiogram showing the recording of normal heart sounds, together with corresponding electrocardiogram tracing. $S_1$ is the first heart sound and marks the beginning of systole. Source [Springer \emph{et al}., 2016].}
\label{fig:fhs}
\end{figure}

PhysioNet \cite{PhysioNet} has held a Computing in Cardiology Challenge since 2000 that requires participants to automatically analyze physiologic time series data. The 2016 challenge \cite{clifford2016classification} asked participants to perform automated analysis of phonocardiogram (PCG) waveforms, i.e. heart sound data collected using digital stethoscopes. The objective of the challenge was to accurately classify normal and abnormal heart sounds. Recordings were collected from both healthy individuals, as well as those with heart disease, including heart valve disease and coronary artery disease. A PCG plot showing the recording of the (normal) sounds made by the heart is given in Figure \ref{fig:fhs}.

Heart disease remains the leading cause of death globally, resulting in more people dying every year due to  cardiovascular disease compared to any other cause of death \cite{web/who/cvd}. Successful automated PCG analysis can serve as a useful diagnostic tool to help determine whether an individual should be referred on for expert diagnosis, particularly in areas where access to clinicians and medical care is limited. 

%In this work, we present an algorithm that accepts PCG waveforms as input and computes heat maps of the time-frequency distribution of signal energy. A deep convolutional neural network is used to automatically classify normal versus abnormal heart sounds. Heart sound segmentation is first performed using a logistic regression hidden semi-Markov model. Spectrograms (heat maps) consisting of 6 cepstral coefficients that capture Mel-frequencies varying over time are derived for overlapping sliding window segments. The convolutional neural network was trained to perform automatic feature extraction and distinguish between normal and abnormal heat maps. Heart sound recordings of variable length are dealt with by computing an ensemble of logit scores for all overlapping segments contained within a recording and maximizing the average scores computed for both abnormal and normal classes. Section \ref{sec:approach} provides further details for each stage of the approach.

In this work, we present an algorithm that accepts PCG waveforms as input and uses a deep convolutional neural network architecture to classify inputs as either normal or abnormal using the following steps:

\begin{description}
\item[1. Segmentation of time series]  A logistic regression hidden semi-Markov model is used to segment incoming heart sound instances into shorter segments beginning at the start of each heartbeat, i.e. the $S_1$ heart sound.
\item[2. Transformation of segments into heat maps] Using Mel-frequency cepstral coefficients, one dimensional time series input segments are converted into two-dimensional spectrograms (heat maps) that capture the time-frequency distribution of signal energy.
\item[3. Classification of heat maps using a deep neural network] A convolutional neural network is trained to perform automatic feature extraction and distinguish between normal and abnormal heat maps.
\end{description}

The contributions of this work are as follows:

\begin{enumerate}
\item We introduce a deep convolutional neural network architecture designed to automatically analyze physiologic time series data for the purposes of identifying abnormalities in heart sounds.
\item Given the cost-sensitive nature of misclassification, we describe a novel loss function used to train the above network that directly optimizes the sensitivity and specificity trade-off.
\item We present results from the 2016 PhysioNet Computing in Cardiology Challenge where we evaluated our algorithm and achieved a Top 10 place finish out of 48 teams who submitted a total of 348 entries.
\end{enumerate}

The remainder of this paper is organized as follows. In Section \ref{sec:relatedwork}, we discuss related works, including historical approaches to automated heart sound analysis and deep learning approaches that process physiologic time series input data. Section \ref{sec:approach} introduces our approach and details each step of the algorithm. Section \ref{sec:sesp} further describes the modified cost-sensitive loss function used to trade-off the sensitivity and specificity of the network's predictions, followed by Section \ref{sec:training}, which details the network training decisions and parameters. Section \ref{sec:results} presents results from the 2016 PhysioNet Computing in Cardiology Challenge and in Section \ref{sec:discussion} we provide a final discussion and end with conclusions in Section \ref{sec:conclusion}.

\section{Related Work}
\label{sec:relatedwork}

Before the 2016 PhysioNet Computing in Cardiology Challenge there were no existing approaches (to the authors' knowledge) that applied the tools and techniques of \emph{``deep learning''} to the automated analysis of heart sounds \cite{liu2016open}. Previous approaches relied upon a combination of feature extraction routines input into classic supervised machine learning classifiers. Features extracted from heart cycles in the time and frequency domains, as well as wavelet features, time-frequency and complexity-based features were input into artificial neural networks \cite{de2007automated,uuguz2012adaptive,uuguz2012biomedical,sepehri2008computerized,bhatikar2005classifier} and support vector machines \cite{maglogiannis2009support,ari2010detection,zheng2015novel} for classification. Previous works have also employed Hidden Markov Models for both segmenting PCG signals into the fundamental heart sounds \cite{springer2014support,springer2016logistic}, as well as classifying normal and abnormal instances \cite{wang2007phonocardiographic,DBLP:journals/eaai/Saracoglu12}.

While there have been many previous efforts applied to automated heart sound analysis, gauging the success of historical approaches has been somewhat difficult, due to differences in dataset quality, number of recordings available for training and testing algorithms, recorded signal lengths and the environment in which data was collected (e.g. clinical vs. non-clinical settings). Moreover, some existing works have not performed appropriate train-test data splits and have reported results on training or validation data, which is highly likely to produce optimistic results due to overfitting \cite{liu2016open}. In this work, we report results from the 2016 PhysioNet Computing in Cardiology Challenge, which evaluated entries on a large \emph{hidden} test-set that was not made publicly available. To reduce overfitting, no recordings from the same subject were included in both the training and the test set and a variety of both \emph{clean} and \emph{noisy} PCG recordings, which exhibited very poor signal quality, were included to encourage the development of accurate and robust algorithms.

The work presented in this paper, is one of the first attempts at applying \emph{deep learning} to the task of heart sound data analysis. However, there have been recent efforts to apply deep learning approaches to other types of physiological time series analysis tasks. An early work that applied deep learning to the domain of psychophysiology is described in \cite{DBLP:journals/cim/MartinezBY13}. They advocate the use of \emph{preference deep learning} for recognizing affect from physiological inputs such as \emph{skin conductance} and \emph{blood volume pulse} within a game-based user study. The authors argue against the use of manual ad-hoc feature extraction and selection in affective modeling, as this limits the creativity of attribute design to the researcher. One difference between the work of \cite{DBLP:journals/cim/MartinezBY13} and ours is that they perform an initial unsupervised pre-training step using stacked convolutional denoising auto-encoders, whereas our network does not require this step and is instead trained in a supervised fashion \emph{end-to-end}.

Similar deep learning efforts that process physiologic time series have also been applied to the problems of epileptic seizure prediction \cite{mirowski2008comparing} and human activity recognition \cite{DBLP:conf/ijcai/HammerlaHP16}.

\begin{figure*}
\includegraphics[width=2.0\columnwidth]{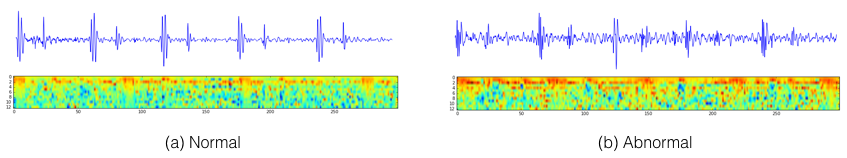}
\caption{MFCC heat map visualization of a 3-second segment of heart sound data. Sliding windows, $i$, are represented on the horizontal axis and filterbank frequencies, $j$, are stacked along the inverted y-axis. MFCC energy information, $c_{i,j}$ is represented by pixel color in the spectrograms. Also shown are the original one-dimensional PCG waveforms that produced each heat map.}
\label{fig:heatmap}
\end{figure*}

\section{Approach}
\label{sec:approach}

Recall from Section \ref{sec:introduction} that our approach consists of three general steps: \emph{segmentation}, \emph{transformation} and \emph{classification}. Each is described in detail below.

\subsection{Segmentation of time series}

The main goal of segmentation is to ensure that incoming time series inputs are appropriately aligned before attempting to perform classification. We first segment the incoming heart sound instances into shorter segments and locate the beginning of each heartbeat, i.e. the $S_1$ heart sound. A logistic regression hidden semi-Markov model \cite{springer2016logistic} is used to predict the most likely sequence of heart sound states ($S_1$  $\rightarrow$ \emph{Systole} $ \rightarrow$ $S_2$ $ \rightarrow$ \emph{Diastole}) by incorporating information about expected state durations. 

Once the $S_1$ heart sound has been identified, a time segment of length, $T$, is extracted. Segment extraction can either be overlapping or non-overlapping. Our final model used a segment length of, $T = 3$ seconds, and we chose to use overlapping segments as this led to performance improvements during initial training and validation.

\subsection{Transformation of segments into heat maps}

Each segment is transformed from a one-dimensional time series signal into a two-dimensional heat map that captures the time-frequency distribution of signal energy. We chose to use Mel Frequency Cepstral Coefficents \cite{davis1980comparison} to perform this transformation, as MFCCs capture features from audio data that more closely resembles how human beings perceive loudness and pitch. MFCCs are commonly used as a feature type in automatic speech recognition \cite{godino2004automatic}. 

We apply the following steps to achieve the transformation:

\begin{enumerate}

\item Given an input segment of length, $T$, and sampling rate, $\nu$, select a window length, $\ell$, and step size, $\Delta$, and extract overlapping sliding windows, $s_i(n)$, from the input time series segment, where $i \in [1, \left \lfloor{\frac{T}{\Delta}}\right \rfloor] $ is the window index and $n \in [1, \ell\nu]$ is the sample index. We chose a window length of 0.025 seconds and a step size of 0.01 seconds.

\item Compute the Discrete Fourier transform for each window.

\begin{equation}
S_i(k) = \sum\limits_{n=1}^{\ell\nu}{s_i(n)h(n)e^{\-i2\pi n \frac{k}{\ell\nu}}}
\end{equation}

where $k \in [1, K]$, $K$ is the length of the DFT and $h(n)$ is a hamming window of length $N$. The power spectral estimate for window, $i$, is then given by (\ref{eqn:power}).

\begin{equation}
\label{eqn:power}
P_i(k) = \frac{1}{N}|S_i(k)|^2
\end{equation}

\item Apply a filterbank of, $j \in [1, J]$, triangular band-pass filters, $d_{j,1 \ldots K}$, to the power spectral estimates, $P_i(k)$, and sum the energies in each filter together. Include a log transformation as sound volume is not perceived on a linear scale.

\begin{equation}
c^*_{i,j} = \log(\sum\limits_{k=1}^{K}d_{j,k}P_i(k))
\end{equation}

We used a filterbank consisting of $J = 26$ filters, where frequency ranges were derived using the Mel scale that maps actual measured frequencies, $f$, to values that better match how humans perceive pitch, $M(f) = 1125 \ln (1 + \frac{f}{700})$.

\item Finally, apply a Discrete Cosine Transform to decorrelate the log filterbank energies, which are correlated due to overlapping windows in the Mel filterbank.

\begin{equation}
c_{i,j} = \sum\limits_{j=1}^{J}{c^*_{i,j}}\cos\Big[\frac{k(2i - 1)\pi}{2J}\Big], k = 1 \ldots J
\end{equation}

The result is a collection of cepstral coefficients, $c_{i,j}$ for window, $i$. For $i = 1 \ldots \left \lfloor{\frac{T}{\Delta}}\right \rfloor$, $c_{i,j}$ can be stacked together to give a time-frequency heat map that captures changes in signal energy over heart sound segments. Figure \ref{fig:heatmap} illustrates two example heat maps (one derived from a normal heart sound input and the other from an abnormal input), where $c_{i,j}$ is the MFCC value (represented by color) at location, $i$, on the horizontal axis and, $j$, on the (inverted) vertical axis.

\end{enumerate}

\subsection{Classification of heat maps using a deep neural network}
\label{sec:architecture}

\begin{figure*}[ht]
%\centering
\includegraphics[width=2.0\columnwidth]{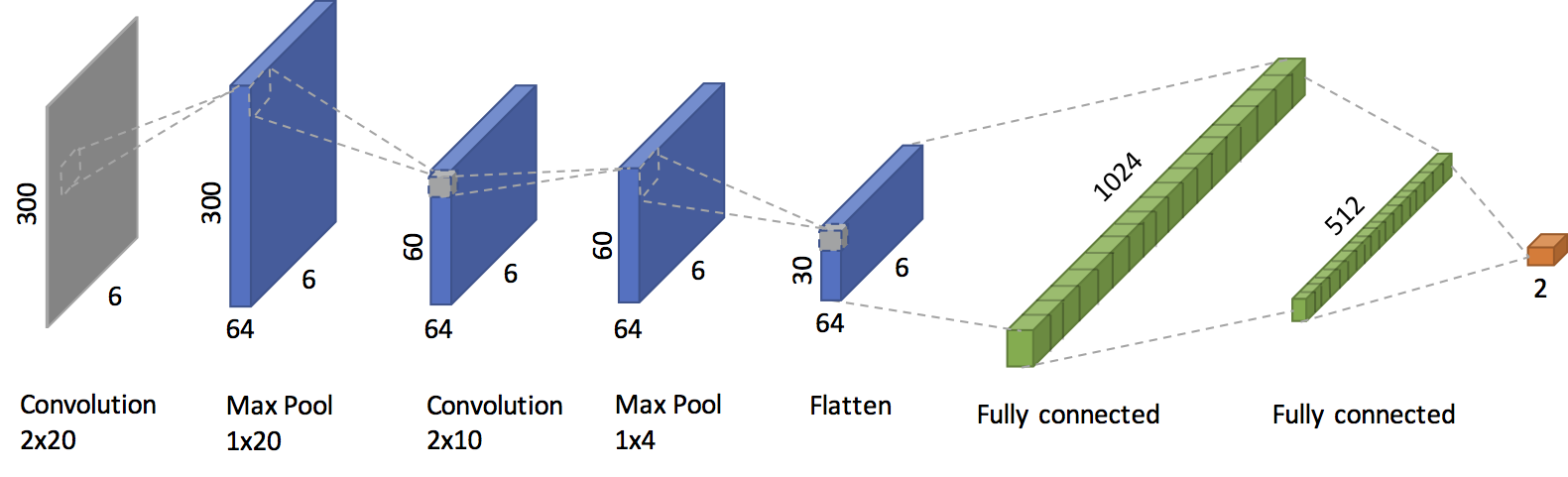}
\caption{Convolutional neural network architecture for predicting normal versus abnormal heart sounds using MFCC heat maps as input}
\label{cnn}
\end{figure*}

The result of transforming the original one-dimensional time-series into a two-dimensional time-frequency representation is that now each heart sound segment can be processed as an image, where energy values over time can be visualized as a heat map (see Figure \ref{fig:heatmap}). Convolutional neural networks are a natural choice for training image classifiers, given their ability to automatically learn appropriate convolutional filters. Therefore, we chose to train a convolutional neural network architecture using heat maps as inputs. 

Decisions about the number of filters to apply and their sizes, as well as how many layers and their types to include in the network were made by a combination of initial manual exploration by the authors, followed by employing a \emph{random search} over a limited range of network architecture parameters. Figure \ref{cnn} depicts the network architecture of a convolutional neural network that accepts as input a single channel 6x300 MFCC heat map and outputs a binary classification, predicting whether the input segment represents a normal or abnormal heart sound. 

The first convolutional layer learns 64 2x20 kernels, using same-padding. This is followed by applying a 1x20 max-pooling filter, using a horizontal stride of 5, which has the effect of reducing each of the 64 feature maps to a dimension of 6x60. A second convolutional layer applies 64 2x10 kernels over the previous layer, once again using same padding. This is again followed by a max-pooling operation using a filter size of 1x4 and a stride of 2, further reducing each feature map to a dimension of 6x30. At this stage in the architecture a flattening operation is applied that unrolls each of the 64 6x30 feature maps into a single dimensional vector of size 11,520. This feature vector is fed into a first fully connected layer consisting of 1024 hidden units, followed by a second layer of 512 hidden units and finally a binary classification output. 

\section{Sensitivity-Specificity Loss Trade-off}
\label{sec:sesp}

The loss function of the network was altered from a standard softmax cross entropy loss function to instead directly trade-off between sensitivity and specificity.

Given unnormalized log-probabilities, $y = Wx + b$, from a classifier consisting of weight matrix, $W$, and bias $b$. The softmax function:

\begin{equation}
s(y_i) = \frac{e^{y_i}}{\sum\limits_j{e^{y_j}}}
\end{equation}

\noindent{gives} probability predictions $P(y_i | x; W, b)$ for the class at index, $i$, for input $x$.

\noindent{Consider,}

\[
Y =
\left[
  \begin{array}{ccc}
    \horzbar & s(y^{(1)}) & \horzbar \\
    \horzbar & s(y^{(2)}) & \horzbar \\
             & \vdots    &          \\
    \horzbar & s(y^{(n)}) & \horzbar
  \end{array}
\right]
,
Y^* =
\left[
  \begin{array}{ccc}
    \horzbar & y^{*(1)} & \horzbar \\
    \horzbar & y^{*(2)} & \horzbar \\
             & \vdots    &          \\
    \horzbar & y^{*(n)} & \horzbar
  \end{array}
\right]
\]

\noindent{where} $s(y_i^{(j)})$, refers to the $i$th entry of row $j$ and $Y^*$ is the corresponding one hot encoded matrix of \emph{actual} class labels.

For the binary class labels of normal ($y_0^*$) and abnormal ($y_1^*$), we define the \emph{mask} matrices, $Y_{Nn}$ and $Y_{Aa}$, where entries within each matrix are softmax prediction values extracted $\forall_{s(y^{(j)}) \in Y}$, as follows:

\[
Y_{Nn} = 
\begin{dcases*}
    s(y_0^{(j)}), & where $y_0^{*(j)} = 1$ and \\
     & $\argmax{\{s(y^{(j)})\}} = \argmax\{y^{*(j)}\} $ \\
      & \\
    0, & otherwise
\end{dcases*}
\]

\[
Y_{Aa} = 
\begin{dcases*}
    s(y_1^{(j)}), & where $y_1^{*(j)} = 1$ and \\
     & $\argmax{\{s(y^{(j)})\}} = \argmax\{y^{*(j)}\} $ \\
      & \\
    0, & otherwise
\end{dcases*}
\]

We then define softmax sensitivity, $S_e$, and specificity, $S_p$, as follows:

\begin{equation}
S_e = \sum\limits_j{\frac{Y_{Aa}^{(j)}}{Y_{Aa}^{*(j)}}},  \hspace{1cm} S_p = \sum\limits_j{\frac{Y_{Nn}^{(j)}}{Y_{Nn}^{*(j)}}}
\end{equation}

\noindent~The final loss function we wish to minimize is given in (\ref{eqn:sesp}).

\begin{equation}
\label{eqn:sesp}
L_{SeSp} = - (S_e + S_p) + \lambda R(W)
\end{equation}

\noindent{where $\lambda R(W)$ is a regularization parameter and routine, respectively.}

\section{Network Training}
\label{sec:training}

$L_2$ regularization was computed for each of the fully connected layers' weight and bias matrices and applied to the loss function. Dropout was applied within both fully connected layers. Table \ref{tab:hyperparameters} shows the values of hyper-parameters chosen by performing a \emph{random search} through parameter space, as well as a list of other network training choices, including weight updates and use of regularization. Adam optimization \cite{DBLP:journals/corr/KingmaB14} was used to perform weight updates. Models were trained on a single NVIDIA GPU with between 4 -- 6 GB of memory. A mini-batch size of 256 was selected to satisfy the memory constraints of the GPU.

\begin{table}
  \centering
  \begin{tabular}{|l|l|}
    \hline
	\bf{Hyper-parameters} & \bf{Value}\\
	\hline
     	Learning rate & 0.00015822\\
     	\hline
	Beta & 0.000076253698849\\
     	\hline
	Dropout & 0.85565561\\	
     	\hline
	\bf{Network parameters} & \bf{Value}\\
	\hline
     	Regularization Type & $L_2$\\
     	\hline
	Batch Size & 256\\
     	\hline
	Weight Update & Adam Optimization\\		
    \hline
	\end{tabular}
  \caption{Listing of hyper-parameters and selected network parameters. Hyper-parameters were learned over the network architecture described in Section \ref{sec:architecture}, using \emph{random search} over a restricted parameter space.}
  \label{tab:hyperparameters}
\end{table}

\subsection{Training/Validation/Test Datasets}

The overall dataset used within the PhysioNet Computing in Cardiology Challenge was provided by the challenge organizers and consisted of eight heart sound databases collected from seven countries over a period of more than a decade \cite{clifford2016classification}. In total 4,430 recordings were taken from 1,072 subjects, resulting in 30 hours of heart sound recordings. From this total dataset, 1,277 heart sound recordings from 308 subjects were removed to be used as held-out test data for evaluating challenge submissions. The test dataset was not made publicly available and challengers were only allowed to make 15 submissions, in total, to the challenge server to evaluate their models on a small 20\% subset of the hidden dataset, before final results were computed. The number of allowed submissions was limited to avoid the issue of participants implicitly overfitting their models on the hidden test dataset. 
%The challenge organizers further ensured that no recordings from the same subject existed in both the training and held-out test dataset.

From the 3153 publicly available PCG waveforms supplied by the challenge organizers, the authors set aside a further 301 instances to be used as a \emph{local held-out test-set} to gauge model performance before making a submission to the challenge server. The remaining instances were used to train initial models. Models were trained on the overlapping 3-second MFCC segments extracted from the remaining 2852 PCG waveforms. This resulted in approximately 90,000 MFCC heat maps, which were split into a training ($\sim75,000$ instances) and validation set ($\sim15,000$ instances). This training and validation set was unbalanced, consisting of approximately 80\% normal segments and 20\% abnormal segments. Training was performed on the unbalanced dataset and no attempt was made to compensate for this class imbalance.

Given that each model was trained on 3-second MFCC heat map segments, it was necessary to \emph{stitch} together a collection of predictions to classify a single full instance. The simple strategy of averaging each class's prediction probability was employed and the class with the greatest probability was selected as the final prediction.

%The authors used the 301 instances, that were initially removed, as a \emph{local held-out test-set} to evaluate a trained model's predictions on full instances, before making a submission to the PhysioNet challenge server. The 301 \emph{local held-out test-set} was a balanced dataset, consisting of approximately 50\% normal and 50\% abnormal instances. 

%Final model evaluation was performed on the challenge server using a completely separate unseen test set.

\section{Results}
\label{sec:results}

\begin{table*}
  \centering
  \begin{tabular}{lcccc}
    \hline
    Rank & Sensitivity & Specificity & Overall & Description\\
    \hline
    1 &	0.9424	&	0.7781 &	0.8602	& AdaBoost \& CNN\\
    2 &	0.8691	&	0.849   &	0.859	& Ensemble of SVMs\\
    3 &	0.8743	&	0.8297 &	0.852	& Regularized Neural Networks\\
    4 & 	0.8639	&	0.8269 &	0.8454	& MFCCs, Wavelets, Tensors \& kNN\\
    5 & 	0.8848	&	0.8048 &	0.8448	& Random Forest + LogitBoost\\
    6 & 	0.8063	&	0.8766 &	0.8415	& Unofficial entry\\
    7 &	0.7696	&	0.9125 &	0.8411	& Probability-distribution based\\
    \bf{8} &	\bf{0.7278}	&	\bf{0.9521} &	\bf{0.8399}	& \bf{Our Approach (see Section \ref{sec:approach})}\\
    9 &	0.8691	&	0.7873 &	0.8282    	& Approach Unknown\\
    10 &	0.7696	&	0.8831 &	0.8263	& Approach Unknown\\
    \hline
    43 &	0.6545  &	0.7569 &	0.7057		& Provided Benchmark Entry\\
    48 &	0.8063	&	0.2643 &	0.5353	& Approach Unknown\\
    \hline    
  \end{tabular}
  \caption{Selected results from the 2016 PhysioNet Computing in Cardiology Challenge}
  \label{tab:results}
\end{table*}

Equations (\ref{eqn:se}) and (\ref{eqn:sp}) show the modified sensitivity and specificity scoring metrics that were used to assess the submitted entries to the 2016 PhysioNet Computing in Cardiology Challenge \cite{clifford2016classification}. Uppercase symbols reflect the true class label, which could either be ($A$)bnormal, or ($N$)ormal. Lowercase symbols refer to a classifier's predicted output where, once again, $a$ is abnormal, $n$ is normal and $q$ is a prediction of unsure. A subscript of 1 (e.g. $Aa_1$, $Na_1$) refers to heart sound instances that were considered good signal quality by the challenge organizers and a subscript of 2 (e.g. $An_2$, $Nn_2$) refers to heart sound instances that were considered poor signal quality by challenge organizers. Finally, the weights used to calculate sensitivity, $wa_1$ and $wa_2$, capture the percentages of good signal quality and poor signal quality recordings in all abnormal recordings. Correspondingly for specificity, the weights $wn_1$ and $wn_2$ are the proportion of good signal quality and poor signal quality recordings in all normal recordings. Overall, scores are given by $\frac{Se + Sp}{2}$.

\begin{equation}
Se = \frac{wa_1 \cdot Aa_1}{Aa_1 + Aq_1 + An_1} + \frac{wa_2 \cdot (Aa_2 +Aq_2)}{Aa_2 + Aq_2 + An_2},
\label{eqn:se}
\end{equation}

\begin{equation}
Sp = \frac{wn_1 \cdot Nn_1}{Na_1 +Nq_1 +Nn_1} + \frac{wn_2 \cdot (Nn_2 + Nq_2)}{Na_2 + Nq_2 + Nn_2}
\label{eqn:sp}
\end{equation}

Table \ref{tab:results} shows a selected subset of the results for the 2016 PhysioNet Computing in Cardiology Challenge. For each selected entry, sensitivity, specificity and overall scores are shown, as well as the entry's final ranking and a brief description of its approach. In total, 348 entries were submitted by 48 teams. Our entry, as described by the algorithm presented in this paper, was ranked 8th with a sensitivity of 0.7278 and specificity of 0.9521, giving an overall score of 0.8399. The top entry to the competition achieved sensitivity of 0.9424, specificity of 0.7781 for an overall score of 0.8602. Also included in Table \ref{tab:results} is the result of a benchmark entry that was supplied by the challenge organizers, which ranked 43rd overall, with a sensitivity of 0.6545 and specificity of 0.7569, for an overall score of 0.7057.

\section{Discussion}
\label{sec:discussion}

Table \ref{tab:results} shows that the overall scores for the top entries to the PhysioNet Computing in Cardiology challenge were very close. In particular, our entry, which achieved an 8th place ranking, had a difference in score of only 0.02, compared to the top place finisher. For our entry, the overall score of 0.8399 was achieved using a single convolutional neural network, whereas other top place finishers achieved strong classification accuracies using an ensemble of classifiers. Improvements in performance have often been witnessed using an ensemble of networks or separate classifiers and we leave this for future work/improvement. For practical purposes, a diagnostic tool that relies on only a single network, as opposed to a large ensemble, has the advantage of limiting the amount of computational resources required for classification. Deployment of such a diagnostic tool on platforms that impose restricted computational budgets, e.g mobile-based, could perhaps benefit from such a trade-off between accuracy and computational cost.

Another point of interest is that our entry to the PhysioNet Computing in Cardiology challenge achieved the greatest specificity score (0.9521) out of all challenge entries. However, the network architecture produced a lower sensitivity score (0.7278). Once again, considering the practical result of deploying a diagnostic tool that relied upon our algorithm, this would likely result in a system with few false positives, but at the expense of misclassifying some abnormal instances. Final decisions about the trade-off between sensitivity and specificity would require further consideration of the exact conditions and context of the deployment environment.

A final point of discussion and area of future improvement is that the approach presented was limited to binary decision outputs, i.e. either normal or abnormal heart sounds. An architecture that also considered signal quality as an output would likely result in performance improvement. 

\section{Conclusion}
\label{sec:conclusion}

The work presented here is one of the first to apply deep convolutional neural networks to the task of automated heart sound classification for recognizing normal and abnormal heart sounds.
We have presented a novel algorithm that combines a CNN architecture with MFCC heat maps that capture the time-frequency distribution of signal energy.
The network was trained to automatically distinguish between normal and abnormal heat map inputs and it was designed to optimize a loss function that directly considers the trade-off between sensitivity and specificity.
We evaluated the approach by submitting our algorithm as an entry to the 2016 PhysioNet Computing in Cardiology Challenge.
The challenge required the creation of accurate and robust algorithms that could deal with heart sounds that exhibit very poor signal quality.
Overall, our entry to the challenge achieved a Top-10 place finish out of 48 teams who submitted 348 entries. 
Moreover, using just a single CNN, our algorithm differed by a score of at most 0.02 compared to other top place finishers, all of which used an ensemble approach of some kind.

\bibliographystyle{named}
\bibliography{ijcai17}

\end{document}